\documentclass[]{spie}  

\usepackage{amsmath,amsfonts,amssymb}
\usepackage{siunitx}
\usepackage{graphicx}
\usepackage{subcaption}
\usepackage{textcomp}
\usepackage[colorlinks=true, allcolors=blue]{hyperref}

\title{Deep ultraviolet resonant Raman (DUVRR) spectroscopy for spectroscopic evaluation and disinfection of food and agricultural samples}

\author[a,*]{Joseph T. Harrington}
\author[b]{Vsevolod Cheburkanov}
\author[b]{Mykyta Kizilov}
\author[b]{Ilya Kulagin}
\author[b]{Georgi Petrov}
\author[a,b]{Vladislav V. Yakovlev}

\affil[a]{Department of Physics and Astronomy, Texas A\&M University, College Station, TX, 77843, USA}
\affil[b]{Department of Biomedical Engineering, Texas A\&M University, College Station, TX, 77843, USA}

\authorinfo{Further author information: (Send correspondence to JTH)\\E-mail: jtharrington@tamu.edu, Telephone: 1 620 255 0667\\}

\begin{document} 
\maketitle

\begin{abstract}
The increasing demands on modern plant and food production due to climate change, regulatory pressures, and the Sustainable Development Goals necessitate advanced photonic technologies for improved sustainability. Deep ultraviolet resonant Raman (DUVRR) spectroscopy offers precise spectral fingerprinting and potential disinfection capabilities, making it a promising tool for agricultural and food sciences. We developed a cost-effective, portable DUVRR spectroscopy system using a mercury (Hg) lamp as the excitation source at 253.65 \unit{\nano\meter}. The system was tested on diverse samples, including alcohol solvents, organic extracts, potential contaminants, and industrial chemicals. The DUVRR system successfully resolved sub-1000 \unit{\per\centi\meter} Raman peaks, enabling detailed spectral fingerprints of various constituents and biomarkers. The system's high sensitivity and specificity ensure precise identification of nutritional values and food quality. The DUV light used in the system, defined here as less than 260 \unit{\nano\meter}, demonstrated potential disinfection properties, adding significant value for food safety applications. The highly sensitive detection capability of our DUVRR system at low powers has significant implications for plant and agricultural sciences. The detailed spectral information enhances the evaluation of nutritional values, food quality, and ripening processes. Moreover, the DUV light's disinfection potential can reduce pathogen load on food surfaces, ensuring safety throughout the food chain. This dually-functional system is highly valuable for precision farming, food production, and quality control. Our DUVRR spectroscopy system provides a highly sensitive, affordable, and portable method for the spectroscopic evaluation and disinfection of food and agricultural samples. Its ability to resolve detailed Raman peaks below 1000 \unit{\per\centi\meter}, combined with DUV light's disinfection capabilities, makes it a promising tool for advancing sustainability and safety in agriculture and food production.
\end{abstract}


\section{INTRODUCTION}
\label{sec:intro}

In an era of rapid population growth and climate change, the agricultural industry is tasked with balancing a delicate equation: producing more food with fewer resources while safeguarding human health and the environment. The pressure to enhance efficiency and maintain sustainability has accelerated the need for innovative solutions in agricultural science. At the same time, the rising global emphasis on food safety and quality demands rapid, accurate, and accessible analytical tools that can detect contaminants, monitor quality, and optimize production processes.\cite{kong_raman_2015,zhang_volume-enhanced_2019,fu_sers-based_2016}

Despite their reliability, conventional analytical methods such as gas chromatography (GC), mass spectrometry (MS), and Fourier-transform infrared (FTIR) spectroscopy are constrained by high costs, complex workflows, and the inability to operate effectively in real-time field conditions. These limitations have been well-documented, particularly in the context of baseline noise, fluorescence interference, and sample preparation challenges. \cite{baek_baseline_2015,ye_baseline_2020,whitaker_simple_2018} Such drawbacks highlight the critical need for a portable, high-resolution technology capable of delivering precise molecular analysis at the point of need.

Deep ultraviolet resonant Raman (DUVRR) spectroscopy has emerged as a revolutionary tool uniquely suited to meet the demands of modern agriculture. Utilizing excitation wavelengths in the deep UV range (e.g., 253.65 \unit{\nano\meter}), this technique achieves a level of precision unmatched by traditional methods.\cite{sterzi_tunable_2019} Unlike visible and near-infrared Raman systems, which are often confounded by fluorescence interference from complex organic samples, DUVRR eliminates this barrier, delivering precise and reliable spectral data even in challenging matrices such as food, biological samples, and agricultural chemicals.\cite{kumamoto_deep_2012,nelson_uv_1992} Additionally, recent advancements have made DUV Raman systems more cost-effective and field-deployable, broadening their practical applications.\cite{troyanova-wood_simple_2013,shutov_highly_2019,harrington2025duvrr}

DUVRR spectroscopy is set apart by its dual capability. The system excels in molecular fingerprinting and incorporates DUV light, which is renowned for its disinfection properties. This functionality addresses a growing concern in the agricultural sector: the control of pathogens and microbial contamination. For instance, studies have demonstrated DUV Raman's ability to detect pathogens with unprecedented sensitivity, making it invaluable for food safety and quality control.\cite{zhang_volume-enhanced_2019,petrov_comparison_2007} By enabling simultaneous analysis and sterilization, DUVRR spectroscopy represents a holistic approach to agricultural challenges.

Furthermore, the system's versatility and portability make it a powerful asset for a variety of agricultural applications. From real-time monitoring of crop health and soil conditions to detecting contaminants in food products and verifying packaging material integrity, this technology redefines the possibilities of on-site analysis. Research shows that portable Raman systems can empower producers, regulators, and researchers alike to make data-driven decisions, enhance productivity, and reduce waste.\cite{egan_one-mirror_2020,kong_raman_2015}

In this paper, we introduce a cost-effective, field-deployable DUVRR spectroscopy system specifically engineered for the agricultural industry. Through rigorous testing across a diverse range of samples, including alcohol solvents, organic extracts, potential contaminants, and industrial chemicals, we highlight the system’s unparalleled performance and practical applications. By addressing the multifaceted challenges of modern agriculture, our system sets a new benchmark for analytical precision and operational efficiency in the industry.

\section{SYSTEM DESIGN AND ADVANTAGES}
\label{sec:methods}

\begin{figure}[h]
    \centering
    \includegraphics[width=0.475\linewidth]{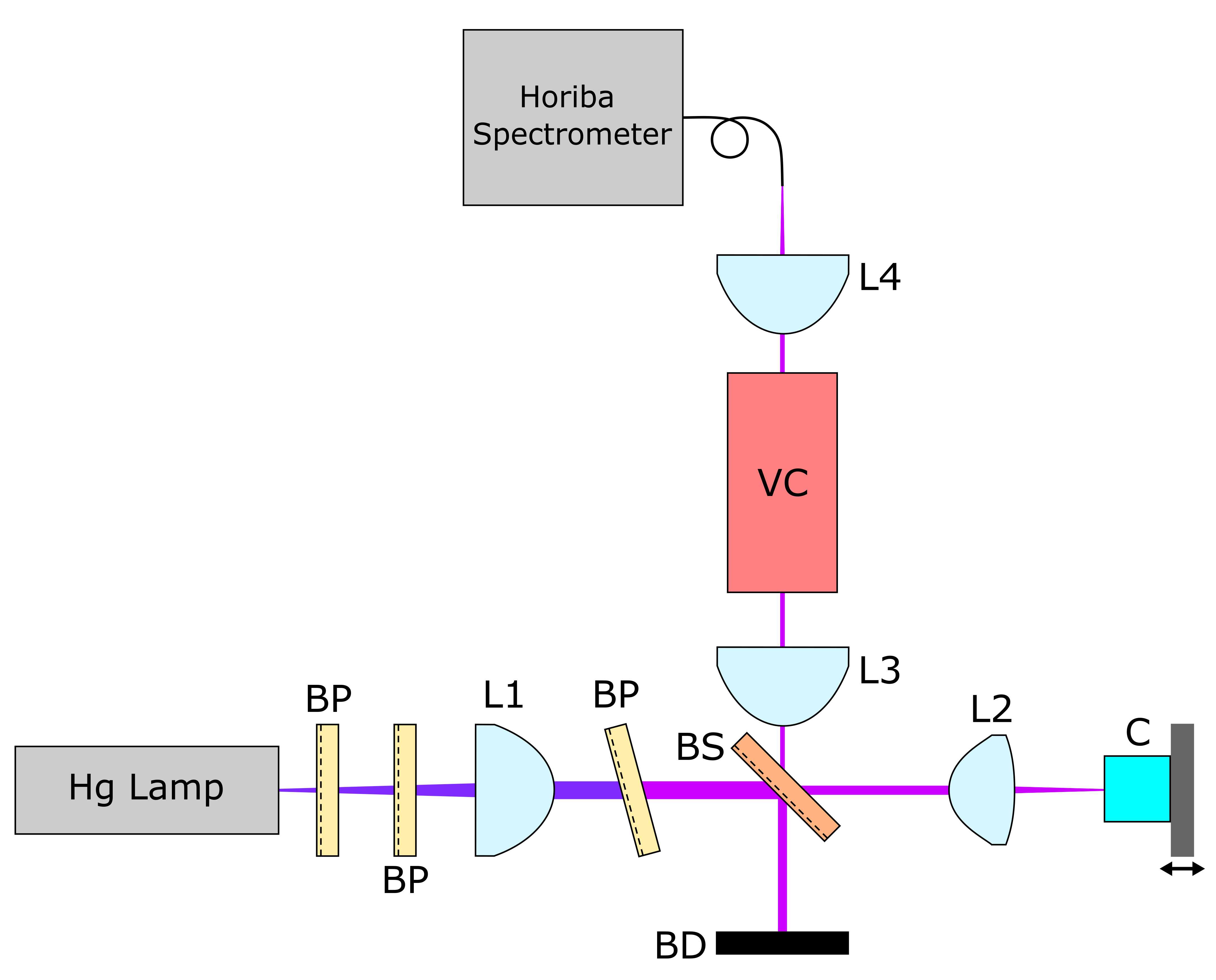}
    \caption{Schematic of the deep ultraviolet resonant Raman (DUVRR) spectroscopy system showcasing the critical components: a mercury (Hg) lamp for excitation at 253.65 \unit{\nano\meter}, bandpass filters (BP) for spectral cleanup, UV aspheric lenses (L1-L4) for collimation and focusing, a beamsplitter (BS) for directional control of light, a mercury vapor cell (VC) for Rayleigh scattering suppression, and a high-resolution spectrometer for detecting Raman spectra.}
    \label{fig:setup}
\end{figure}

\subsection{Light Source}

For our system, we explored several low-cost excitation sources for DUVRR spectroscopy.  Initially, we employed a spectral calibration lamp (6035 Spectral Calibration Lamp; Newport, Inc.). While its nominal lifetime of 5000 hours made it an attractive option for initial alignment and preliminary characterization, the low emitted power (approximately 5 \unit{\nano\watt} after spatial filtering and collimation) limited its practicality for most spectroscopic applications. This led us to select a low-pressure capillary lamp (CPG-1; Jelight Company, Inc.). This lamp, providing over 1 \unit{\micro\watt} of collimated power at 253.65 \unit{\nano\meter}, has proven effective in our setup. It has reliably operated for over three years without replacement or noticeable power degradation, serving as the primary excitation source for the reported measurements. Unlike traditional laboratory systems, the lamp’s durability and operational efficiency are well-suited for on-site agricultural applications. This enables in-the-field monitoring of soil, crops, and agricultural inputs, ensuring data-driven decision-making.

\subsection{Raman Filtering}

Choosing a Hg emission line as the excitation source enabled us to leverage the same Hg vapor for Raman signal filtering. Conventional Raman edge or notch filters often fail to adequately suppress elastic scattering in low-frequency Raman shifts ($<1000$ \unit{\per\centi\meter}) applications. Using a Hg vapor cell, we effectively suppressed Rayleigh scattering, enhancing the signal-to-noise ratio for critical Raman features. This capability is particularly valuable for analyzing complex agricultural samples, such as soil nutrient profiles or plant biochemical markers. Accurate molecular fingerprinting ensures precise nutrient management and early detection of plant stressors. Moreover, recent work has demonstrated the capability of hyperspectral Raman imaging systems to resolve complex molecular signatures in a single shot when there is a high signal-to-noise ratio.\cite{thompson_single-shot_2017}

\subsection{Optical Layout}

The experimental setup for our DUVRR spectroscopy system is meticulously designed to optimize both performance and ease of use. The diagram in Fig.~\ref{fig:setup} illustrates the complete system configuration, highlighting the interplay between various components.

The low-pressure mercury lamp emits at 253.65 \unit{\nano\meter} and is filtered using a series of 254 \unit{\nano\meter} MaxLamp™ mercury line filters (Semrock, Inc.) and Schott UG5 glass. This combination effectively mitigates extra emission lines of mercury. The cleaned-up light is collimated using a deep ultraviolet (DUV) fused silica aspheric lens (Thorlabs, Inc.). It then passes through a rotated 266 \unit{\nano\meter} MaxLine™ laser line bandpass filter (Semrock, Inc.). The filter is rotated to "blue shift" the bandpass for additional spectral cleanup and power attenuation. A UV fused silica plate (Thorlabs, Inc.) is used as a beamsplitter, transmitting half of the light to the sample. The remaining light is focused onto the sample using a high numerical aperture (NA = 0.63) UV anti-reflection (AR)-coated aspheric lens (Edmund Optics, Inc.), which doubles as the collection optic. Collected light propagates back through the beamsplitter, where half is reflected toward the mercury vapor cell (VC). A DUV fused silica aspheric lens (Thorlabs, Inc.) ensures the light is properly coupled into the vapor cell, suppressing Rayleigh scattering by several orders of magnitude. The remaining Raman signal is coupled into a fiber bundle via another aspheric lens and directed to an iHR 320 Horiba spectrometer. The spectrometer is equipped with a 2400 grooves/\unit{\milli\meter} grating (330 \unit{\nano\meter} blaze) optimized for DUV and a liquid nitrogen-cooled charge-coupled device (CCD) for long acquisition times.

This setup ensures optimal suppression of undesired scattering, high signal collection efficiency, and precise spectral resolution. Each component is carefully selected to optimize performance in the DUV spectral range.

\subsection{Key Advantages}

\begin{itemize}
    \item \textbf{Fluorescence Suppression:} Fluorescence from complex organic samples is a major limitation of visible or near-infrared Raman spectroscopy. Deep UV excitation circumvents this issue, offering superior spectral clarity of organic agricultural materials, such as food products and plant extracts.
    \item \textbf{Dual Functionality:} The inclusion of DUV light facilitates molecular identification and provides a means of surface disinfection. This feature is particularly valuable for reducing microbial contamination on crops and food products.
    \item \textbf{Portability and Cost-effectiveness:} The compact design and use of readily available components make the system ideal for deployment in field settings. Unlike bulky laboratory instruments, our system offers analysis at a fraction of the cost.
    \item \textbf{Versatility:} The system’s ability to analyze various sample types—including liquids, solids, and biological matrices—makes it a powerful tool for diverse agricultural applications.
\end{itemize}

\section{RESULTS AND DISCUSSION}
\label{sec:results}

The experiments conducted with the DUVRR spectroscopy system highlight its capability to provide detailed molecular fingerprints across a range of samples. The system integrates precise excitation sources, effective Raman filtering, and high-sensitivity detection components, ensuring consistent, reproducible results, even for complex samples. Key findings from the analysis of alcohol solvents, organic extracts, potential contaminants, and industrial chemicals showcase the versatility and accuracy of this system.

\subsection{Alcohol Solvents Analysis}

\begin{figure}[h]
    \centering
    \begin{subfigure}{0.475\linewidth}
    \includegraphics[width=\linewidth]{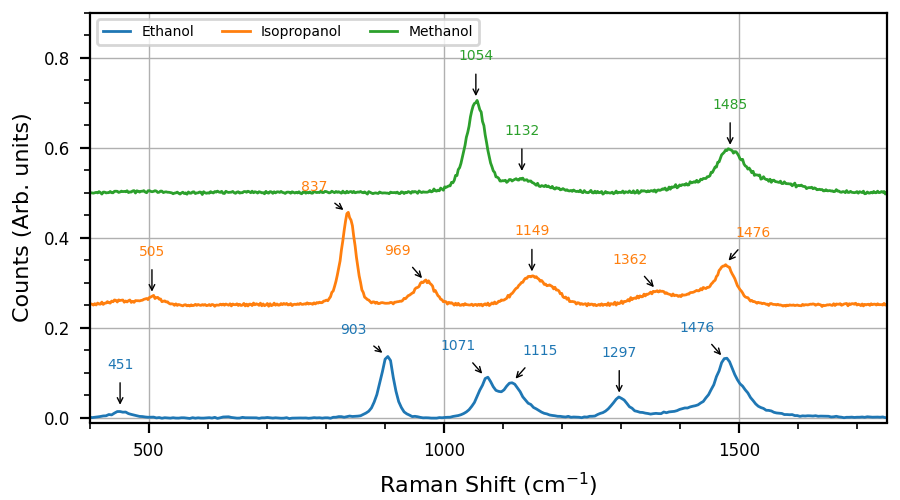}
    \caption{}
    \label{fig:alcohols_1}
    \end{subfigure}
    \hfill
    \begin{subfigure}{0.475\linewidth}
    \includegraphics[width=\linewidth]{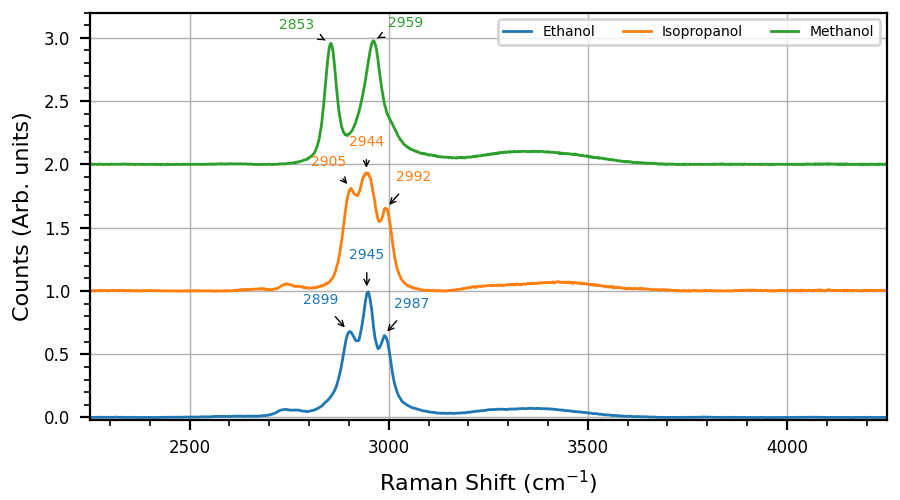}
    \caption{}
    \label{fig:alcohols_2}
    \end{subfigure}
    \caption{Raman spectra of various alcohol solvents obtained using the DUVRR spectroscopy system, showcasing distinct molecular fingerprints in two spectral ranges: (a) 400–1750 \unit{\per\centi\meter}, which includes vibrational modes of functional groups, and (b) 2250–4250 \unit{\per\centi\meter}, corresponding to C-H and O-H stretching vibrations. These spectra demonstrate the system's capability for precise molecular identification in industrial applications such as alcohol quality control.}
\end{figure}

The DUVRR spectroscopy system demonstrated its efficacy in analyzing alcohol solvents by providing distinct molecular fingerprints. In Figs.~\ref{fig:alcohols_1} and \ref{fig:alcohols_2}, we show spectral ranges between 400–1750 \unit{\per\centi\meter} revealed unique vibrational modes of functional groups, while the 2250–4250 \unit{\per\centi\meter} range captured C-H and O-H stretching vibrations. These features allowed for precise molecular identification, confirming the system's suitability for applications like alcohol-based disinfectant quality control. The high specificity and reproducibility of these results indicate a significant advantage over conventional spectroscopic techniques, particularly in field-based industrial applications.

\subsection{Organic Extracts Analysis}

\begin{figure}[h]
    \centering
    \includegraphics[width=0.475\linewidth]{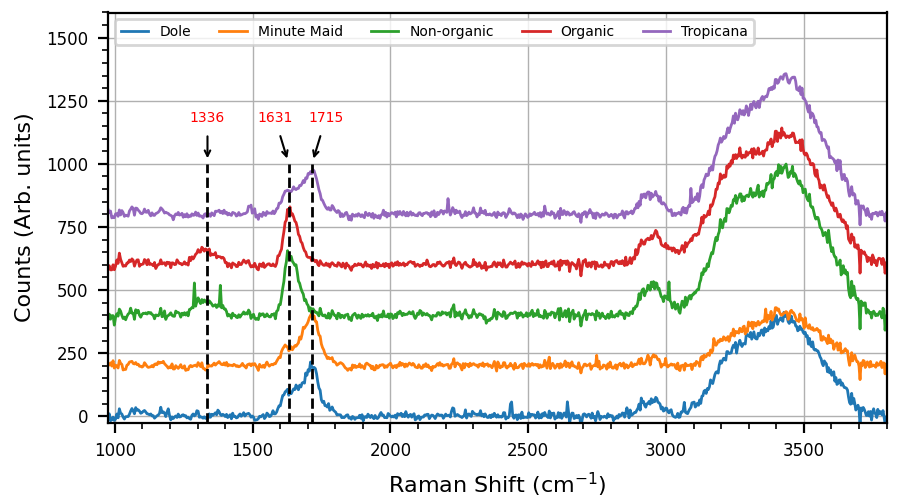}
    \caption{Comparison of Raman spectra of organic extracts from raw apple juice (organic and non-organic) and commercially processed apple juice, revealing molecular differences due to processing and compositional variation. The data highlights the system's ability to distinguish nutritional and chemical properties, supporting food quality assessment and authenticity verification applications.}
    \label{fig:juices}
\end{figure}

The evaluation of organic extracts, including raw and processed apple juices like that in Fig.~\ref{fig:juices}, revealed molecular differences attributable to processing and compositional variations. For instance, the system distinguished raw from processed samples by identifying key biochemical markers. This capability underscores the potential for DUVRR spectroscopy in nutritional assessment, food quality verification, and authenticity checks. The sensitivity of the system to subtle molecular variations highlights its value for ensuring regulatory compliance and consumer trust.

\subsection{Detection of Contaminants}

\begin{figure}[h]
    \centering
    \includegraphics[width=0.475\linewidth]{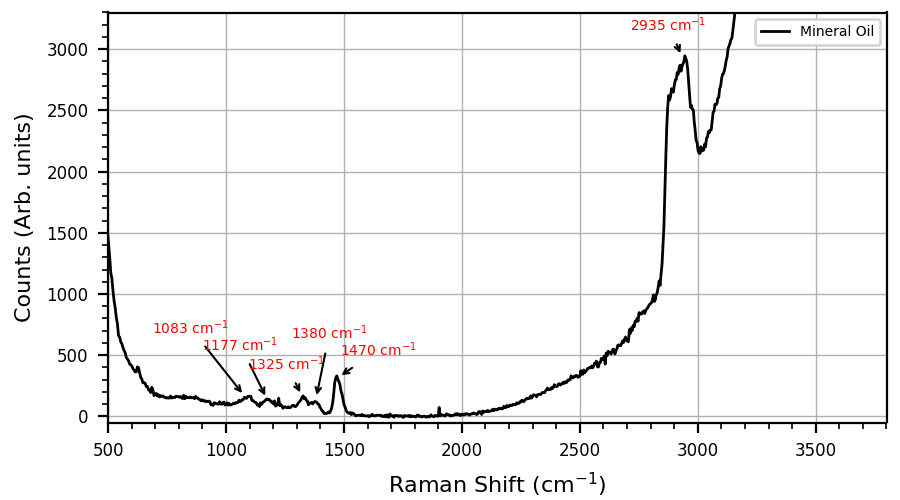}
    \caption{Raman spectrum of mineral oil showing characteristic molecular fingerprints in the DUV range. This spectrum emphasize the system’s sensitivity and specificity for identifying trace impurities in complex food and industrial samples.}
    \label{fig:mineral_oil}
\end{figure}

The system's high sensitivity facilitated the detection of trace contaminants such as mineral oil, shown in Fig.~\ref{fig:mineral_oil}. Characteristic molecular fingerprints in the DUV range were identified, emphasizing the system's role in enhancing food safety. The ability to detect low concentrations of impurities positions DUVRR spectroscopy as a critical tool for quality control in food production and industrial settings.

\subsection{Agriculture Applications}

\begin{figure}[h]
    \centering
    \begin{subfigure}{0.475\linewidth}
    \includegraphics[width=\linewidth]{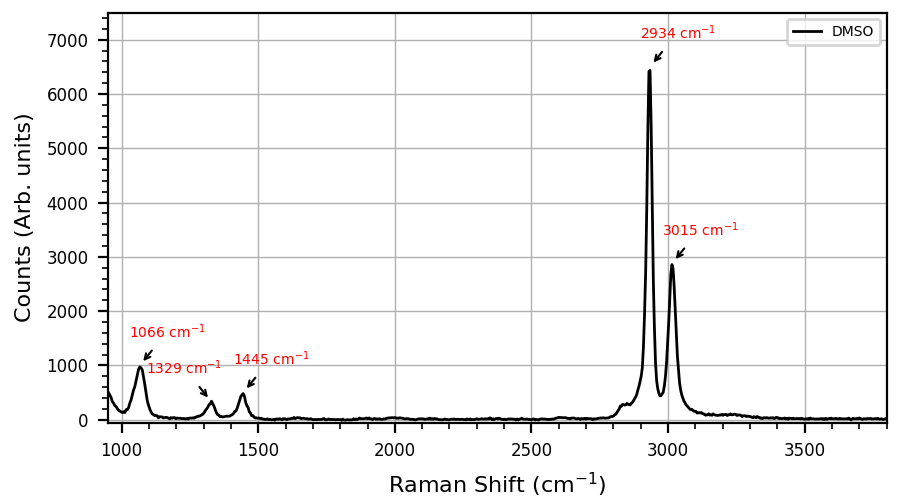}
    \caption{}
    \label{fig:dmso}
    \end{subfigure}
    \hfill
    \begin{subfigure}{0.475\linewidth}
    \includegraphics[width=\linewidth]{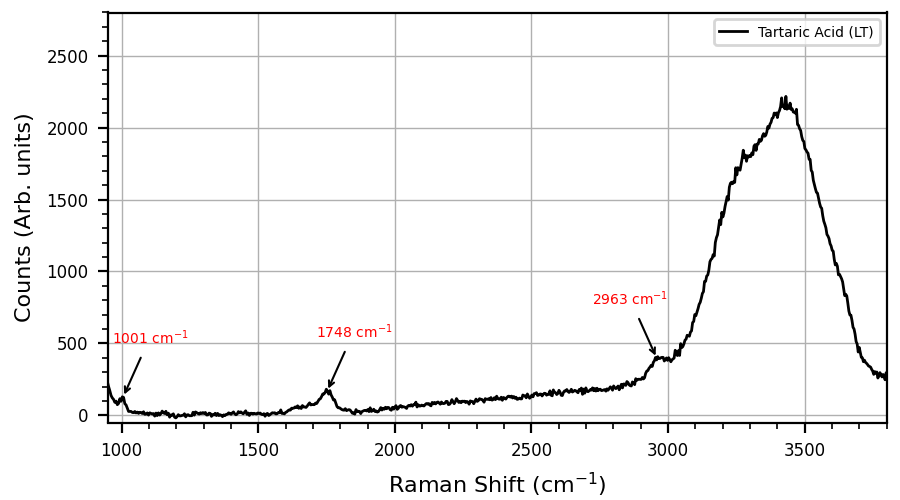}
    \caption{}
    \label{fig:tartaric_lt}
    \end{subfigure}
    \caption{Raman spectra of agriculturally relevant compounds: (a) dimethyl sulfoxide (DMSO), often used as a solvent, and (b) tartaric acid (LT), a key organic acid in agricultural and food applications. These results demonstrate the system’s ability to resolve complex molecular structures, aiding in chemical monitoring and agricultural research.}
\end{figure}

The analysis extended to agriculturally relevant compounds, including dimethyl sulfoxide (DMSO) and tartaric acid, as seen in Figs.~\ref{fig:dmso} and \ref{fig:tartaric_lt}. These results demonstrated the system’s capacity for resolving complex molecular structures, aiding in monitoring agricultural chemicals and soil nutrients. 

\subsection{Discussion of Implications}

\subsubsection{Advancements in Agricultural Science}

The results highlight the untapped potential of DUVRR spectroscopy in agricultural science:

\begin{itemize}
    \item \textbf{Precision Farming:} The ability to monitor soil and crop health on-site enables data-driven decision-making, optimizing resource use and reducing waste.
    \item \textbf{Food Safety:}The system’s ability to detect contaminants and evaluate nutritional content enhances quality control measures, fostering consumer trust and regulatory adherence.
    \item \textbf{Sustainability:} Portable, cost-effective systems enable widespread adoption, supporting sustainable agricultural practices.
\end{itemize}

Techniques such as the thermostable Raman interaction profiling (TRIP) method have shown remarkable accuracy in estimating amino acid compositions, further underscoring the value of Raman spectroscopy in food quality and safety assessments.\cite{altangerel_novel_2024}

\subsubsection{Broader Industrial Applications}

Beyond agricultural applications, Raman spectroscopy has proven effective in pharmaceutical development, including the precise analysis of protein-ligand interactions, which further emphasizes its versatility across industries including \cite{altangerel_label-free_2023}:

\begin{itemize}
    \item \textbf{Pharmaceuticals:} The precise identification of molecular structures is critical for quality assurance in drug manufacturing.
    \item \textbf{Environmental Monitoring:} The system’s sensitivity to trace compounds facilitates the detection of pollutants in water and soil.
    \item \textbf{Polymers and Manufacturing:} Our system is well-suited to compose structural integrity analysis of polymers to support compliance with industrial standards, ensuring product reliability.
\end{itemize}

\section{CONCLUSION AND FUTURE PERSPECTIVES}
\label{sec:conclusion}

Our DUVRR spectroscopy system represents a transformative approach for the agricultural industry. Combining high-resolution molecular analysis with DUV disinfection capabilities, this portable and cost-effective tool addresses key challenges in food safety, quality assurance, and sustainability. The system’s versatility and performance make it an invaluable asset for advancing agricultural science and ensuring the integrity of the global food supply chain.

The findings underscore the need for further development and integration of DUVRR spectroscopy. Potential advancements include:

\begin{itemize}
    \item \textbf{Integration with IoT:} Real-time data sharing and analysis could enable predictive diagnostics in agricultural and industrial contexts.
    \item \textbf{AI-Driven Insights:} Machine learning algorithms could enhance spectral interpretation, improving accuracy and usability.
    \item \textbf{Expanded Applications:} Extending research to diverse sample types and environmental conditions will broaden the technology’s impact.
\end{itemize}

\appendix    

\acknowledgments 
 
JTH received funding support via the SMART Scholarship-for-Service Program. VVY received partial funding support from the Air Force Office of Scientific Research (AFOSR) (grants FA9550-20-1-0366, FA-9555-23-1-0599), the National Institutes of Health (NIH) (grants R01GM127696, R01GM152633, R21GM142107, R21CA269099), and NASA, BARDA, NIH, and USFDA, under Contract/Agreement No. 80ARC023CA002.

\bibliography{report} 
\bibliographystyle{spiebib} 

\end{document}